\RequirePackage{fix-cm}
\documentclass{svjour3}                     
\smartqed  
\usepackage{graphicx}
\usepackage{amssymb}
\begin{document}

\title{Some clarifications on the relation between Bohmian quantum potential and Mach's principle}

\author{Faramarz Rahmani\and Mehdi Golshani}
\institute{F. Rahmani \at
              School of Physics, Institute for Research in Fundamental Science(IPM), Tehran, Iran\\
              Tel.: +98-21-22180692,
              Fax: +98-21-22280415\\
              \email{faramarzrahmani@ipm.ir}           
           \and
           M. Golshani \at
              School of Physics, Institute for Research in Fundamental Science(IPM), Tehran, Iran\\
              Department of Physics, Sharif University of Technology, Tehran, Iran\\
              Tel.:+98-21-66022718, Fax.:+98-21-66022718\\
              \email{golshani@sharif.edu}
}              
\date{Received: date / Accepted: date}

\maketitle

\begin{abstract}
Mach's principle asserts that the inertial mass of a body is related to the distribution of other distant bodies. This means that in the absence of other bodies, a single body has no mass. In this case, talking about motion is not possible, because the detection of motion is possible only relative to other bodies. But in physics we are faced with situations that are not fully Machian. As in the case of general theory of relativity where geodesics exist in the absence of any matter, the motion has meaning. Another example which is the main topic of our discussion, refers to Bohmian quantum mechanics, where the inertial mass of a single particle does not vanish, but is modified. We can call such situations in which motion or mass of a single particle has meaning, pseudo-Machian situations. In this paper, we use the Machian or pseudo-Machian considerations to clarify under what circumstances and how a Machian effect leads us to Bohmian quantum mechanics. Then, we shall get the Bohmian quantum potential and its higher order terms for the Klein-Gordon particle through Machian considerations, without using any quantum mechanical postulate or operator formalism. 
\end{abstract}

\section{Introduction}
\label{intro}
According to the Mach principle, absolute space has no meaning and space is only the distance between bodies. \cite{Refbucket,Refbarbur1} In other words, space without the existence of matter is not definable or understandable, i.e. the existence of space is due to existence of matter. This means that, space is not an independent entity.  Therefore it does not affect the dynamics of matter in Mach's view.\cite{Refbucket,Refbarbur1} \par Furthermore, the inertial mass of an object is related to the existence  of matter in other places. It may be interpreted as action at a distance gravity. Newton's gravity is action at a distance, and in his view, the mass of a particle is a property of the particle. But in Mach's view, the mass is dependent on the distribution of other masses all over the universe. In the Newtonian gravity, the mass of a particle does not depend on other particles, and gravitational force is dependent on the distance between bodies in the real space. But, in a Machian theory, since the mass of a particle is related to other masses, the inverse dependency on distance is not necessary. In fact, the Machian force is a force in the configuration space of the system and it varies with respect to the changes of the configuration of the system. For an example about the nature of Machian force or potential, consider a triangle with three angles $\theta_1$,$\theta_2$ and $\theta_3$; and the sides $a$,$b$ and $c$. If we multiply the sides by a constant factor $\lambda$, we will have a bigger $(\lambda>1)$ or smaller $(\lambda<1)$ triangle with respect to the primary triangle without changes in the angles between the sides and configuration of the system. So, such potential remains invariant under rescaling. Furthermore, when we multiply the distances between particles by a constant factor, the total volume of the system is also multiplied by a constant factor. Now, since the total mass of universe is considered to be constant, we conclude that the density of matter is also multiplied by a constant factor. But all of these do not change the mass of particles. Because there is no change in the configuration of system. The value of this argument will be clear in the next section where we want to obtain the form of the Machian potential. The configuration of matter in the universe is approximately constant at large scales. So we get constant mass for a particle. Now, suppose that the configuration of matter changes rapidly. Then the amount of change in the particle mass will be significant.\par Another statement of Mach's principle is that the motion of a particle, should be considered relative to the other objects. Otherwise, the motion is not detectable or there is no motion in Mach's view at all. As we know from the Newtonian mechanics, the mass of a particle, which is a property of that particle, is defined as the resistance of the particle against the change in its movement. Newton defined the acceleration of an object relative to the absolute space or inertial frames. In Newtonian mechanics, when an external force $\vec{F}$ is exerted on a body, it experiences an acceleration $\vec{a}$ relative to an inertial frame $S $ or other inertial frames that are in uniform motion relative to this frame. The problem arises when we consider a frame $S^{\prime}$ that has the acceleration $\vec{a}^{\prime}$ relative to the frame $S$. In this new frame, Newton's second law becomes:
\begin{equation}\label{iner}
\vec{F}-m\vec{a}^{\prime}= m\vec{a}^{\prime \prime}
\end{equation}
where, $\vec{a}^{\prime \prime}$ is the acceleration of body in $S^{\prime}$. The term $m\vec{a}^{\prime}$
is not related to external forces. In Newton's view it is due to the acceleration of reference frame relative to absolute space. Newton called it "inertial force". Practically, he considered distant stars as frames of absolute space, which are inertial frames. Here, Mach noted that the absolute space of Newton is fictitious and what Newton actually did is considering the motion relative to distant stars or distant bodies and absolute space has no meaning. According to Mach's considerations, if there is no matter in the universe, then for a single particle we have $m\vec{a}=0$. Then since $\vec{a}$ is indeterminate, as there is no objects to define the motion of particle relative to it, thus we conclude that $m=0$. This means that in the absence of other objects a particle has no mass. In other words, the dynamics and the mass of a particle is determined by the whole matter of the universe, even distant parts of universe.\cite{Refbucket} Naturally, in this view, mass is a dynamical quantity not a property of matter. In Newton's own view, since motion takes place relative to an absolute space, so we conclude that only the acceleration of a particle relative to absolute space vanishes, not its mass. From the Newton's point of view, the inertial mass of a body is a property of that body, whether other bodies exist or do not exist. In Mach's view, the mass of a particle is related to mass distribution of matter all over the universe. Thus the mass of a particle can not be considered as a property of the particle. There are different views on space, time and mass. See, e.g., refs. \cite{Refbarbur2,RefMax1,RefMax2} \par If we consider Mach's principle to be true, then another principle of physics i.e weak equivalence principle of general relativity (WEP), which states that the inertial mass and gravitational mass of any object are equal, will be violated. The WEP states that the behavior of a free falling particle is universal. Under the Mach considerations the WEP can not be true. Because, in the general theory of relativity, the inertial mass of a particle is considered to be a property of that particle, and not a dynamical variable, depending on the distribution of other bodies. In general theory of relativity inertial mass is seen like the gravitational mass which is a property of matter. If, we consider the Machian definition of mass, the relation 
\begin{equation}\label{ee}
\vec{a}=\vec{g}
\end{equation}
for a free falling particle can not be true, because this is due to the equivalence of inertial mass and gravitational mass, but in the Machian view, the inertial mass is due to existence of other matter. So relation (\ref{ee}) is not correct. Since, in Mach's view, the mass and consequently the dynamics of a particle depends on other particles, so Machian considerations lead us to a configuration space formalism automatically. Machian effects are nonlocal. This is in conflict with Einstein's view which is based on local interactions of matter.\par 
With all this taken into account, there are some cases that prevent us from accepting the Mach principle comprehensively. That part which refers to the action of all distant mater on a body seems acceptable. However, it may be in conflict with special relativity and the problem of restricted transition of information. The other part which states that in the absence of other matter the mass of a single particle vanishes and also the motion is meaningless is somewhat suspicious. Here, we argue about this aspect of Mach's principle in a different context. \par
Our first example comes from the general theory of relativity. Consider Einstein's equations:
\begin{equation}
R_{\mu \nu}=8\pi G (T_{\mu \nu}-\frac{1}{2}Tg_{\mu \nu})
\end{equation}
If we put $T_{\mu \nu}=0$, then we have $R_{\mu \nu}=0$. This situation is not identical to the flat space-time of special relativity. Because, in the above equation, the Ricci tensor $R_{\mu \nu}$ has 10 independent component, while at the same time the Rieman tensor $R_{\mu\nu\rho\lambda}$, which is the space-time curvature tensor, does not vanish completely. Because it has 20 independent components. Naturally, we can have a geodesics with a moving particle on it without any other matter. What does this geodesics mean? This leads us to believe in space as an independent being, without needing any matter. In other words, space is not only the distances between objects. This brings to mind that there should be an ethereal entity which we can consider motions relative to it. If we consider space as an entity that has energy, then it is a kind of matter. Thus Mach's principle could be retrieved again.\par  Einstein has also referred to a dynamical ether whose qualities, like metric and curvature, determines the dynamics of matter locally in a covariant way. Here, we point to a letter that Einstein wrote to Lorentz in June 1916:\\
\textit{I agree with you that general relativity theory admits of an ether hypothesis as does the special relativity theory. But this new ether theory would not violate the principle of relativity. The reason is that the state [...metric tensor] = Aether is not that of a rigid body in an independent state of motion, but a state of motion which is a function of position determined through the metrical phenomena.}\cite{Refkost}\par
Our aim from referring to this argument is that Mach's relative space hypothesis does not hold completely. In other words, the motion of a single object has meaning relative to this new ether. \par
Now, we present another example from the deterministic and causal version of quantum mechanics which was provided by David Bohm in 1952.\cite{RefBohm}. Bohm's theory of quantum physics is a causal but nonlocal theory. In Bohmian quantum mechanics, the wave-function $\psi$, which is represented in configuration space, has a real origin. It is sometimes called  pilot-wave. Because its main property is guidance of particles on real trajectories. But, since we can not provide the initial conditions exactly and on the other hand there are always noises in the environment, so the final results are obtained statistically \cite{RefHolland}. In Bohm's own view, the quantum behavior of matter is due to an unknown origin whose effects are appear as a non-local quantum potential in the Hamilton-Jacobi equation of system. For example, the quantum Hamilton-Jacobi of a non-relativistic particle becomes:
\begin{equation}\label{ha}
\frac{\partial S(p,q)}{\partial t}+H(p,q)+\left(-\frac{\hbar ^2}{2m}\frac{\nabla^2 \sqrt{\rho(q,t)}}{\sqrt{\rho(q,t)}}\right)=0
\end{equation}
where $\sqrt{\rho(q,t)}$ is the amplitude of pilot-wave $\psi =\sqrt{\rho}e^{\frac{iS}{\hbar}}$. For many-body systems, this amplitude is represented by $\sqrt{\rho(q_1,q_2, ... q_n,t)}$. The last term of the equation (\ref{ha}) is called Bohmian quantum potential. The quantity $S(p,q)$ which is the phase of the pilot-wave which is Hamilton's principal function. In a many-body system, the velocity of ith particle is related to distribution of all other particles:
\begin{equation}
\frac{d \vec{x}_i}{dt}= \frac{\nabla_i S(q_1,...,q_n,t)}{m_i}
\end{equation} 
This induces a Machian behavior in quantum mechanics. Because, the motion of a particle is related to that of other particles instantaneously. We want to clarify how the mass of a single particle has meaning, while there is no matter.\par
In the relativistic version of Bohmian quantum mechanics, for a single spinless particle, we have a meaningful relation for its rest mass as:
\begin{equation}\label{mass}
\mathcal{M} =m_0 \sqrt{1+\frac{\hbar^2}{m_0^2}\frac{\partial_\mu \partial^{\mu} \sqrt{\rho}}{\sqrt{\rho}}}
\end{equation}
where the second term under radical is the quantum potential of a relativistic particle. This is obtained for a single-particle system through the principles of Bohmian mechanics.\cite{RefHolland}. The relativistic extension to the case of many-particle pilot-wave is possible. See for example ref \cite{RefGold} for fermions and \cite{Refnik1,Refnik2} for bosons. But here we confine ourselves to a one-particle problem, because we want to discuss about another topic. According to Mach's principle, the rest mass of a single object should vanishes. But the relation (\ref{mass}) demonstrates that in the quantum domain, for a causal and deterministic theory, there is a type of unavoidable dependency of particle mass on the mass distribution $\rho$ which is related to quantum wave function $\psi$ through the relation $\rho=\mid \psi \mid ^2$, which does not vanish even in single-particle mode. This is itself a Machian property while there is only one particle! The world "pseudo-Machian" may be more suitable for it. If there is no matter, then what does the dependency of single particle mass on $\rho$ mean? In the next section we introduce a suitable definition for the mass distribution function. We shall see that for getting the relation (\ref{mass}), we should define the distribution function relative to an absolute space. So the relation (\ref{mass}) is a pseudo-Machian relation. For obtaining the form of pseudo-Machian potential the existence of an absolute space is necessary. Then we shall examine to see if it is possible to obtain the relation (\ref{mass}) or similar relations through Machian considerations, without using any quantum mechanical postulations? If we can do this, then we have demonstrated that the quantum effects have a Machian or pseudo-Machian nature. In following, we try to represent these relations under particular conditions.

\section{Constructing a Machian-type interaction}
\label{sec:1}
According to Mach's statement, the inertial mass of a particle can be considered as a functional which relates its mass  to the relative mass distribution function $\rho_{\rm{rel}}$ of other particles. So, we write:
\begin{equation}\label{m1}
\mathcal{M}_i = \mathcal{M}_i[\rho_{\rm{rel}}(x^\mu_1(\lambda),x^\mu_2(\lambda),\cdots,\hat{x}^\mu_i(\lambda), \cdots, x^\mu_n(\lambda))]
\end{equation}
The parameter $\lambda$ is an auxiliary scalar parameter that parametrizes particle's trajectories, like the proper time in relativity. The circumflex on symbol $x_i$ means that symbol has been omitted. Because, according to Mach's principle a single body has no mass and $\rho_{\rm{rel}}$ is relative to the ith particle.   
If we empty the universe from any matter and keep only one particle, then there should not be any force present. Therefore, according to Mach's statement the mass of a single particle becomes zero. In other words, $\mathcal{M}_i [\rho_{\rm{rel}}=0]=0$. Also, the motion of the particle will not be detectable. Naturally, such definition does not lead us to a result such relation (\ref{mass}). Therefore, we should use another definition for the mass distribution function.\par Now, we want to investigate another scenario which can embraces the Bohmian-type situations. We consider the mass distribution $\varrho$ of all matter relative to an absolute space. In this case, if we have only one particle, it always has a distribution relative to the absolute space. Then universal time $\lambda$, which we introduced, could refer to the absolute space. This sort of definition may not be a fully Machian statement. Nevertheless, it has a Machian aspect, because the mass of a particle is related to the mass distribution function of the particle itself. This distribution function is defined relative to an absolute space. We can call such model a pseudo-Machian model. Of course, if we consider the space as a background of matter and energy, then the theory turns into a fully-Machian theory. Our discussion does not contain the effects of matter on such absolute space or in general the action-reaction principle. We only consider the effect of absolute space on matter in this formalism. Because the nature of absolute space is not clear to us. We only consider it for necessity. For a more detailed study about the action-reaction principle ref. \cite{Refbrown} is suitable. Since, relation (\ref{m1}) is inefficient for our purpose, we use another interpretation. But, we continue to use the definition:
\begin{equation}\label{m2}
\mathcal{M}_i =\mathcal{M}_i[\varrho(x^\mu_1(\lambda),x^\mu_2(\lambda),\cdots,x^\mu_i(\lambda), \cdots, x^\mu_n(\lambda))]
\end{equation}
The essential difference between relations (\ref{m1}) and (\ref{m2}) is that in relation  (\ref{m2}) the mass of ith particle never vanishes, because $\varrho$ also depends on existence of ith particle. This is because we have considered mass distribution relative to the absolute space. \textit{This can be seen as the effect of absolute space on a physical system}. It may be interpreted as an ethereal behavior. The relation (\ref{m2}) can be stated with more precision. In fact, In addition to the $\varrho$, any spatial or temporal change in the distribution of matter relative to the absolute space affects the mass of the ith particle. Because according to Mach's arguments the mass of a particle should  be related to the distribution of other matter instantaneously. Since, we have considered $\varrho$ with respect to the absolute space, we can have a more precise relation for one-particle system in the form:
\begin{equation}\label{m3}
\mathcal{M} =\mathcal{M}[\varrho(x(\lambda)), \partial_\mu \varrho(x(\lambda)), \partial_\mu \partial_\nu \varrho(x(\lambda)),\cdots]
\end{equation} 
This means that if we have a single particle, the spatial or temporal variations of the particle with respect to absolute space affects the mass of that particle itself. \textit{Naturally, if we do not consider an absolute space, the relation (\ref{m3}) has no meaning for a single particle}. Furthermore, when a particle moves from a point to another point in every instance it gets a new configuration with respect to the absolute space. So in addition to equation (\ref{m3}) for a Machian mass, we could have:
\begin{equation}\label{m4}
\mathcal{M}=\mathcal{M}(x(\lambda))
\end{equation} 
We shall use both definitions (\ref{m3}) and (\ref{m4}) in following. First, we use relation (\ref{m4}) for obtaining a general form for the Machian mass. \par
Now, let us investigate the effect of Machian mass on the geodesics of a particle. The geodesics equation of a particle, with constant rest mass $m_0$, is derived through the variations of the famous action:
\begin{equation}\label{ac1}
\mathcal{A}=\int \frac{1}{2}m_0 g_{\mu\nu}\frac{dx^\mu}{d \lambda}\frac{dx^\nu}{d \lambda}d\lambda
\end{equation}
Now, we use this action for a particle with Machian mass $\mathcal{M}=\mathcal{M}(x(\lambda))$ . For simplicity, we do calculations in a flat space-time, and then we shall convert the result to curved space-time for discussing about WEP. We can write:
\begin{equation}\label{ac2}
\mathcal{A}_{\rm{Machian}}=\int \frac{1}{2}\mathcal{M}(x(\lambda)) \frac{dx^\mu}{d \lambda}\frac{dx_\mu}{d \lambda}d\lambda
\end{equation}
This action has already been introduced. See ref. \cite{Refdick}. For infinitesimal displacements, we impose diffeomorphism: $x^{\mu}\longrightarrow x^{\mu}+\xi^{\mu}$, and from this, we get:
\begin{eqnarray}
\delta \mathcal{A}_{\rm{Machian}}&=&\frac{1}{2}\int \delta \mathcal{M}(x(\lambda)) \frac{dx^\mu}{d\lambda}\frac{dx_\mu}{d\lambda}d\lambda +\frac{1}{2}\int \mathcal{M}(x(\lambda))\delta\left(\frac{dx^\mu}{d\lambda}\frac{dx_\mu}{d\lambda}\right)d\lambda \\ \nonumber &=& \frac{1}{2}\int \partial_\nu \mathcal{M}(x(\lambda))\xi^\nu  \frac{dx^\mu}{d\lambda}\frac{d x_\mu}{d\lambda}d\lambda +\int \mathcal{M}(x(\lambda))\frac{dx_\mu}{d\lambda} \frac{d\xi^\mu}{d\lambda}d\lambda
\end{eqnarray}
Substituting the identity:
\begin{equation}
\frac{d}{d\lambda}\left(\mathcal{M}\frac{dx_\mu}{d\lambda}\xi^\mu \right)= \mathcal{M}\frac{dx_\mu}{d\lambda} \frac{d\xi^ \mu}{d \lambda}+ \frac{d}{d\lambda}\left(\mathcal{M}\frac{dx_\mu}{d\lambda}\right) \xi^\mu
\end{equation}
into the second integral, and considering that the variation of $x^\mu$ at boundaries is zero, we get:
\begin{equation}
\int \left(\frac{1}{2}\partial_\nu \mathcal{M} \frac{dx^\mu}{d\lambda}\frac{dx_\mu}{d\lambda} -\frac{d}{d\lambda}\left(\mathcal{M}\frac{dx_\mu}{d\lambda}\right) \right)\xi^\nu d\lambda =0
\end{equation}
Since this relation holds for an arbitrary region of space-time, we get:
\begin{equation}\label{modgeoflat}
\frac{d^2 x_\mu}{d \lambda ^2}=\frac{1}{2}\left(\eta_{\mu \nu} -\frac{dx_\mu}{d\lambda}\frac{dx_\nu}{d\lambda} \right) \frac{\partial ^\nu \mathcal{M}}{\mathcal{M}}
\end{equation}
 Extension to a curved space-time is easy. Then we get to:
\begin{equation}\label{geocurved}
\frac{d^2 x^ \mu}{d \lambda ^2} + \Gamma ^{\mu} _{\nu\sigma}\frac{dx^\nu}{d\lambda}\frac{dx^ \sigma}{d\lambda}=\frac{1}{2}\left( g^{\mu\nu} - \frac{dx^\mu}{d\lambda}\frac{dx^ \nu}{d\lambda}  \right)\frac{\nabla_\nu \mathcal{M}}{\mathcal{M}}
\end{equation}
This is not a geodesic equation, since the  right-hand side of the above equation is due to Machian effects. The non-relativistic limit of geodesics equation in any textbooks is derived through the 
some considerations. In the non-relativistic limit, the particle velocity is small relative to the light velocity; so $\frac{dx^i}{d\lambda}\ll  \frac{dt}{d\lambda}$. The space-time metric has a small deviation with respect to the metric of flat space-time. 
\begin{equation}
g_{\mu \nu} =\eta_{\mu \nu} + h_{\mu \nu} ,\quad \vert h_{\mu \nu}\vert \ll  1
\end{equation}
Using the inverse metric relation, we have:
\begin{equation}
g^{\mu \nu}g_{\nu \sigma}=\delta ^{\mu} _{\sigma} \Longrightarrow g^{\mu \nu}=\eta ^{\mu \nu} -h^{\mu \nu}
\end{equation} 
where, $h^{\mu \nu} =\eta ^{\mu \rho}\eta ^{ \nu\sigma} h_{\rho \sigma}$. Now, we assume that in the non-relativistic regime the metric is stationary i.e, $\partial_{0} g_{\mu \nu} =0$ and Machian mass has no explicit dependency on time i,e $\partial_{0} \mathcal{M}(x^i)=0$, where $x^i$ stands for spatial coordinates of the particle, with $i=1,2,3$.  The parameter $\lambda$ becomes equal to the parameter $t$ which is time in non-relativistic regime. If we collect all things together, we get:
\begin{equation}\label{ac1}
\frac{d^2 \vec{x}}{dt^2}=\vec{\nabla} (\frac{GM}{r}) -\frac{1}{2} \frac{\vec{\nabla} \mathcal{M}(\vec{x})}{\mathcal{M}(\vec{x})}
\end{equation}
or
\begin{equation}\label{ac2}
\vec{a}=\vec{g} -\frac{1}{2} \frac{\vec{\nabla} \mathcal{M}(\vec{x})}{\mathcal{M}(\vec{x})}
\end{equation}
This equation shows that in a Machian-type formalism, WEP can not be held anymore. In other words, in a local gravitational field, the inertial acceleration will not be equal to the gravitational acceleration. The equation (\ref{ac2}) is comparable with equation (\ref{iner}), which represents a particle in a non-inertial frame with acceleration $\frac{1}{2} \frac{\vec{\nabla} \mathcal{M}(\vec{x})}{\mathcal{M}(\vec{x})}$ relative to an inertial frame. From the Machian point of view, this acceleration is due to the existence of other matter. But as we mentioned before, by using the concept of density function relative to the absolute space, this is also true for a single-particle system. So this is a pseudo-Machian relation. We shall see that one of the consequences of this choice is getting to Bohmian quantum mechanics without postulate wave function from the beginning. \par
For the case which $\vec{g}=0$, we have:
\begin{equation}\label{ac3}
\vec{a}= -\frac{1}{2} \frac{\vec{\nabla} \mathcal{M}(\vec{x})}{\mathcal{M}(\vec{x})}
\end{equation}
By multiplying both sides of the above equation by $m_0$ which is the mass of particle in non-Machian regime, and by integrating we have:
\begin{eqnarray}\label{work}
-(\frac{m_0}{2})\int \frac{\vec{\nabla} \mathcal{M}(\vec{r})\cdot d \vec{r}}{\mathcal{M}(\vec{r})}&=& m_0 \int \vec{a}\cdot d\vec{r}\nonumber\\
-(\frac{m_0}{2})\int \frac{d\mathcal{M}}{\mathcal{M}}&=& W_{\rm{M}}\nonumber \\
\mathcal{M}(\vec{r})&=&\alpha \exp \left(-{\frac{2W_{\rm{M}}}{m_0}}\right)\nonumber \\ &=& \alpha \exp (\frac{2 Q_{nr}}{m_0})
\end{eqnarray}
where, the $W_{\rm{M}}=-Q_{nr}$ is the work which is done by the Machian force. The expression $Q_{nr}$ is the non-relativistic Machian potential and $\alpha$ is a constant with the dimension of mass. For the special case where the Machian acceleration is constant, we have:
\begin{equation}
\mathcal{M}=\alpha e^{2\vec{a}\cdot \vec{r}}
\end{equation}
So, Mach's statement leads us to conclude that the mass of a particle is related to the Machian acceleration of that particle. This is not the case in Newtonian mechanics or in Einstein's relativity. 
We  use the form of relation (\ref{work}) in the next section to suggest a similar relation for the relativistic case.

\section{Machian interaction for a relativistic spinless particle }
\label{sec:2}
The relation (\ref{m2}), shows that configuration space is at the heart of Mach's principle. So, it is suitable to use Hamilton-Jacobi formalism. For a single relativistic spinless particle, the Hamilton-Jacobi equation is:
\begin{equation}\label{classicalhamilton}
\partial_\mu S \partial ^\mu S -m_0^2 =0
\end{equation}
where, $S$ is Hamilton's principal function. This relation can be derived from the following action in the space $(\varrho,S)$ i,e.
\begin{equation}\label{densityhamiltonian}
\mathcal{A}=\int \varrho\left(\partial_\mu S \partial ^\mu S -m_0^2 \right)d^4 x
\end{equation}
Now, by using the relation (\ref{m3}), the Machian action is:
 \begin{eqnarray}\label{amach}
 \mathcal{A}_{\rm{Machian}}&=&\int \varrho \left(\partial _\mu S \partial^{\mu} S - \mathcal{M}^2 \right)d^4 x \nonumber \\
 &=&\int \varrho \left(\partial _\mu S \partial^{\mu} S - \mathcal{F}[\varrho(x), \partial_\mu \varrho(x), \partial_\mu \partial_\nu \varrho(x),\cdots]\right)d^4 x
 \end{eqnarray}
The non-relativistic equation (\ref{work}) suggests a relativistic ansatz in the form: 
\begin{equation}\label{qw}
\mathcal{M}^2=\mathcal{F}=\beta \exp \left(\mathcal{Q}\right)
\end{equation}
Naturally, we should have $\beta=m_0^2$, because when the Machian effects tend to zero, i.e. $\mathcal{Q} \longrightarrow 0$, then equation (\ref{densityhamiltonian}) must be retrieved. The first condition  was the definition of density relative to absolute space. Here, we want to impose a second condition. We postulate that Machian effects have a small correction on the energy of a particle relative to $m_0^2$. Thus, we do not solve a very general equation. Now, from equation (\ref{qw})we have:
\begin{equation}\label{app}
\mathcal{M}^2=\mathcal{F}\approxeq m_0^2(1+\mathcal{Q})\quad , \quad \frac{Q}{m_0^2}\ll 1
\end{equation}
Then, equation (\ref{amach}) becomes:
\begin{equation}\label{aamach}
 \mathcal{A}_{\rm{Machian}}=\int \varrho \left(\partial _\mu S \partial^{\mu} S - m_0^2-\mathfrak{Q}[\varrho(x), \partial_\mu \varrho(x), \partial_\mu \partial_\nu \varrho(x),\cdots]\right)d^4 x
\end{equation}
where $\mathfrak{Q}=m_0^2 \mathcal{Q}$. Then, we should have:
\begin{equation}\label{vari}
\delta_{\varrho} \mathcal{A}_{\rm{Machian}}=\delta \int \varrho \left(\partial _\mu S \partial^{\mu} S - m_0^2-\mathfrak{Q}[\varrho(x), \partial_\mu \varrho(x), \partial_\mu \partial_\nu \varrho(x),\cdots] \right)d^4 x=0
\end{equation} 
Using relation (\ref{vari}) and by variation with respect to $\varrho$, we get:
\begin{equation}\label{38}
\left[\partial_\mu S \partial^\mu S -m^2 -\mathfrak{Q}[\varrho]\right]+ \varrho \frac{\partial \mathfrak{Q}}{\partial \varrho}- \partial_\mu \left(\varrho\frac{\partial \mathfrak{Q}}{\partial(\partial_\mu \varrho)}\right)+ 
\partial_\mu \partial^\mu \left(\varrho \frac{\partial \mathfrak{Q}}{\partial(\partial_\mu \partial^\mu \varrho)}  \right)=0\nonumber
\end{equation}
The expression in the bracket vanishes, because for the new Hamilton-Jacobi equation, like any Hamilton-Jacobi equation, we should have:
\begin{equation}
\partial _\mu S \partial^{\mu} S - m_0^2-\mathfrak{Q}[\varrho(x), \partial_\mu \varrho(x), \partial_\mu \partial_\nu \varrho(x),\cdots]=0
\end{equation}
Thus, from the (\ref{38}) we get:
\begin{equation}\label{motion}
\varrho \frac{\partial \mathfrak{Q}}{\partial \varrho}- \partial_\mu \left(\varrho\frac{\partial \mathfrak{Q}}{\partial(\partial_\mu \varrho)}\right)+\partial_\mu \partial^\mu \left(\varrho \frac{\partial \mathfrak{Q}}{\partial(\partial_\mu \partial^\mu \varrho)}  \right)=0
\end{equation}
Now, our task is to suggest a typical form for $\mathfrak{Q}$, using Machian considerations. We suggest a Machian potential in the form:
\begin{equation}\label{potentials}
\mathfrak{Q}=C ((\varrho)^r)^m (\partial_\mu (\varrho)^r)^n(\partial_\mu \partial^\mu (\varrho)^r)^p
\end{equation}
In following, we articulate about our suggestion. Here, we consider $\varrho ^ r$, instead of $\varrho$ directly. Because we want to have a formula closer to the relation (\ref{mass}). So, we should determine the value of $r$ through the variation of action. 
It is possible to take higher order derivatives in (\ref{potentials}), but we choose a simple form and investigate its consequences. The power $n$ should be even, otherwise, $\mathfrak{Q}$ can not be a scalar quantity. The power $p$ can be even or odd because the term in the parenthesis is always a scalar. For the choice of power $m$, we postulate that a Machian-type potential does not depend on distance, like in real space; rather it is affected by the configuration space of particles or particle. Because, this is not a potential or force due to a local source, like many forces in physics, which decreases or increases with distance. According to the our argument about the Machian potential in the introduction, if we multiply the  density $\varrho$ by a constant factor $\gamma$, the value of the potential should not change. Thus, by imposing this condition on relation (\ref{potentials}), third condition becomes:
\begin{equation}\label{condition}
\mathfrak{Q}[\varrho^{\prime}=\gamma \varrho]=\mathfrak{Q}[\varrho] \Longrightarrow \gamma^{r(m+n+p)}=1,\quad r\neq 0
\end{equation} 
From this condition we obtain a helpful condition for selecting powers in relation (\ref{potentials}):
\begin{equation}\label{condition2}
m+n+p=0,
\end{equation}
The simplest choice is $m=-1$,\quad $n=0$\quad and\quad $p=1$. Naturally, $m$ should be negative integer. Otherwise, it can not cancel the constant factors of other terms in (\ref{potentials}) after multiplying $\varrho$ by a constant factor. Other choices are possible, but this is the simplest choice. In general, we can write such potentials in the Taylor expansion form that has been studied for the non-relativistic case in references \cite{RefAtigh,RefAtigh2}. Using this choice, we have:
\begin{equation}\label{potential2}
\mathfrak{Q}=C\varrho^{-r} \partial_\mu \partial^\mu \varrho^r
\end{equation}
After expanding this relation, we get:
\begin{equation}\label{potential3}
\mathfrak{Q}=C\left(r(r-1)\frac{(\partial_\mu \varrho)^2}{\varrho^2}+r\frac{\partial_\mu \partial^\mu \varrho}{\varrho}\right)
\end{equation}
Every term in the parenthesis satisfies condition (\ref{condition2}) separately. 
By substituting equation (\ref{potential3}) into the equation (\ref{motion}), and after some calculations, we get:
\begin{equation}
\frac{\partial_\mu \partial^\mu\varrho}{\varrho}+ 2(r-1)\frac{\partial_\mu \partial^\mu \varrho}{\varrho}=0 \Longrightarrow r=\frac{1}{2}
\end{equation}
This is an interesting result. Because, this shows that the power $\frac{1}{2}$ in relation (\ref{mass}) is not arbitrary rather it is due to a particular condition.
Now, if we collect all things together, we conclude that the Machian potential for a spinless relativistic particle is:
\begin{equation}
\mathfrak{{Q}}=C \frac{\partial ^\mu \partial _\mu \sqrt{\varrho}}{\sqrt{\varrho}}
\end{equation}
or 
\begin{equation}\label{lam}
\mathcal{Q}=\frac{\mathfrak{Q}}{m^2_0}=\frac{C}{m_0^2} \frac{\partial ^\mu \partial _\mu \sqrt{\varrho}}{\sqrt{\varrho}}
\end{equation}
This is surprising. Because, if we put $C=\hbar^2$ we get the Bohmian quantum potential of a spinless particle exactly, in simplest form. Unfortunately, we do not have any criterion to find where we can put $C=\hbar^2$ in relation(\ref{lam}), rather we use quantum mechanics for obtaining constant $C$. From Bohmian quantum mechanics, we know that $\sqrt{\varrho}=R$, where $R$ is the amplitude of pilot-wave which has been defined from the beginning in Bohm's own formalism. Here we obtained $r=\frac{1}{2}$ and its consequent $\sqrt{\varrho}=R$ automatically, without using quantum mechanics. We did not want to obtain only the Bohmian quantum potential of a relativistic spinless particle; rather, our aim was to clarify the relation between Machian statements and Bohmian quantum potential. By using the relations (\ref{app}) and (\ref{lam}), we get:
\begin{equation}\label{77}
\mathcal{M}^2 =m^2_0 \left(1+\frac{\hbar^2}{m^2_0}\frac{\Box \sqrt{\varrho}}{\sqrt{\varrho}}\right)
\end{equation}
which is the square of the relation (\ref{mass}). This demonstrates that under special conditions the quantum effects have Machian or pseudo-Machian origins. On the other hand, the definition of a Machian mass field enters an unavoidable acceleration. This shows if we impose three aforementioned conditions, that quantum effects could be related to the problem of accelerated frames. It may be said that since the above equation was derived for a single-particle system, thus it is not a Machian relation. But it is evident that the mass $\mathcal{M}$ is related to $\varrho$ and its derivatives. This means that the mass of a particle is not an inherent property. Since, this has been obtained through the definition of an absolute space. We can say that absolute space has causal effects on physical systems. We can attribute an agent $\psi$ to represent the effects of absolute space on physical system. We can write $\varrho=R^2$ as $\varrho=Re^{(if)}e^{(-if)}R$, where from quantum mechanics we know that $f=\frac{S}{h}$ and $\varrho=\psi^{*} \psi$, where $\psi$ is the same as wave function or the pilot wave in Bohmian quantum mechanics which was postulated there from the beginning. But, we obtained this results without using any quantum mechanical postulates and wave function. This argument demonstrated that what we consider as an absolute space has quantum mechanical effects on physical systems under special conditions. It is wiser that do not consider the absolute space as an empty space. Rather it is a kind of ether with mysterious properties. But the nature of such ether is not clear to us.\par 
Now, we want to obtain the next term for the Machian potential. This is not possible in Bohm's original approach. For simplicity, we work with $R=\sqrt{\varrho}$ in following. According to relations (\ref{qw})and (\ref{lam}), the third term of relation (\ref{qw}) after expansion is proportional to $\frac{(\Box R)^2}{R^2}$, which is related to the powers $m=-2$, $n=0$ and $p=2$ in the condition (\ref{condition2}). This term $\frac{(\Box R)^2}{R^2}$ can be written in another form which we investigate it in the following. Since, $r=\frac{1}{2}$, thus, the equation (\ref{potentials}) takes the form:
\begin{equation}\label{new}
\mathfrak{Q}= C(R)^m (\partial_\mu R)^n(\partial_\mu \partial^\mu R)^p
\end{equation} 
Now, the variational equation (\ref{motion}) becomes: 
\begin{equation}\label{2var}
R^2 \frac{\partial \mathfrak{Q}}{\partial R}- \partial_\mu \left(R^2\frac{\partial \mathfrak{Q}}{\partial(\partial_\mu R)}\right)+\partial_\mu \partial^\mu \left(R^2 \frac{\partial \mathfrak{Q}}{\partial(\partial_\mu \partial^\mu R)}  \right)=0
\end{equation}
Now, we substitute the relation (\ref{new}) into relation (\ref{2var}), and after some straightforward but long calculations we find that for satisfying (\ref{2var}), when $m=-2$, $n=0$ and $p=2$, we should have:
\begin{equation}
(\partial_\mu \partial^\mu R)^2=R (\partial_\mu \partial^\mu \partial_\mu \partial^\mu R)
\end{equation}
If we use $\Box^2$ instead of $\partial_\mu \partial^\mu \partial_\mu \partial^\mu$ schematically, we get
\begin{equation}
(\Box R)^2=R (\Box^2 R)
\end{equation}
Thus, a more precise relation for (\ref{qw}) becomes:
\begin{equation}
\mathcal{M}^2 =m^2_0 \left(1+\frac{\hbar^2}{m^2_0}\frac{\Box \sqrt{\varrho}}{\sqrt{\varrho}}+\frac{1}{2}\frac{\hbar^4}{m^4_0}\frac{\Box^2 \sqrt{\varrho}}{\sqrt{\varrho}}+\cdots\right)
\end{equation}
Machian considerations allow us to obtain higher order terms for the Bohmian quantum potential. We can say that Bohmian quantum potential, through the standard approach of Bohm, is a special type of Machian or pseudo-Machian potential.
Here, we do not use any quantum mechanical postulates and operator formalism. Also, we obtained higher order degrees terms for the Bohmian potential.

\section{conclusion}
\label{sec:4}
In Bohmian quantum mechanics, quantum potential is responsible for quantum mechanical energies. We found that the quantum potential comes from a Machian or pseudo-Machian theory under special conditions. We called this approach a pseudo-Machian approach. We obtained this result through the three conditions. First, we should define the distribution mass function with respect to an absolute space. With this assumption, we can consider Machian mass for a single particle system. Second, the energies of Machian effects are very small relative to the rest mass of the particle. Third, the form of Machian potential does not change if we multiplying the density by a constant factor. With the three aforementioned assumptions, we could relate a Machian or pseudo-Machian potential to the quantum potential of a relativistic spinless particle. On the other hand, the Machian mass leads us to an unavoidable acceleration which is like the acceleration of a frame with respect to an inertial frame. This shows the close relation between the quantum mechanics, accelerated frames and the absolute space under the three aforementioned conditions. Here, we obtained quantum mechanics from Machian considerations and the difference between quantum world and classical world is in the coefficient $e^Q$. An important point is that we obtained Bohmian results from the perspective of Machian statements, without postulating any quantum mechanical assumptions, like the postulation of a wave function from the beginning. Also, we found that the wave function is not an original concept. Rather, it is an auxiliary instrument which relates the effects of the unknown absolute space on matter, through an unknown process. We demonstrated that the derivation of  Bohmian potentials through the Machian considerations is possible without using the concept of wave function. Finally, the Machian approach allows us to obtain higher order terms for Bohmian quantum potential.
\begin{acknowledgement}
This work was supported in part by the National Elite Foundation of Islamic Republic
Of Iran.
\end{acknowledgement}



\end{document}